\documentclass[aps,prd,twocolumn,amsmath,longbibliography,amssymb,nofootinbib]{revtex4-2}

\usepackage[T1]{fontenc}
\usepackage{lmodern}
\usepackage{amsfonts}
\usepackage{bm}
\usepackage{graphicx}
\usepackage{booktabs}
\usepackage{microtype}
\usepackage{hyperref}
\usepackage{xcolor}

\definecolor{linkblack}{RGB}{20,20,20}
\definecolor{citeblue}{RGB}{0,70,140}
\definecolor{urlblue}{RGB}{0,90,120}

\hypersetup{
  colorlinks=true,
  linkcolor=linkblack,
  citecolor=citeblue,
  urlcolor=urlblue
}

\allowdisplaybreaks
\begin{document}

\title{Untwisting the double copy: the zeroth copy as an optical seed}

\author{Damien~A.~Easson}
\email{easson@asu.edu}
\affiliation{Department of Physics, Arizona State University, Tempe, Arizona 85287-1504, USA}

\author{Michael~J.~Falato}
\email{mjfalato@asu.edu}
\affiliation{Department of Physics, Arizona State University, Tempe, Arizona 85287-1504, USA}

\begin{abstract}
We present a historical optical foundation for stationary vacuum Kerr--Schild spacetimes on a flat background and interpret it in modern double-copy language. In this setting, a complex optical seed \(\rho=-\theta-i\omega\), built from the expansion and signed twist of the Kerr--Schild congruence, is harmonic, while its inverse obeys an eikonal equation and reconstructs the congruence algebraically. Thus the local stationary geometry is organized by a single complex seed. In the overlap of the stationary Kerr--Schild and Petrov type--D Weyl double-copy framework, this seed furnishes a normalized representative of the zeroth-copy data, while its real part yields the Kerr--Schild profile and its gradient generates the single-copy gauge-field strength. The construction provides, without recourse to twistor methods, a spacetime realization of how a single complex seed builds the congruence, organizes the associated spacetime and gauge fields, and encodes the geometric content of the zeroth copy.
\end{abstract}

\maketitle

\section{Introduction}
\label{sec:intro}

The classical double copy relates exact gravitational solutions to gauge and scalar data, providing a spacetime realization of structures first uncovered in scattering amplitudes \cite{Kawai:1985xq,Bern:2008qj,Bern:2010ue}; see also \cite{White:2017mwc}. The original Kerr--Schild double-copy paradigm applies to stationary Kerr--Schild metrics on a flat background and identifies a corresponding gauge-field single copy and scalar profile---often interpreted retrospectively as a zeroth copy---on Minkowski spacetime~\cite{Monteiro:2014cda}. Subsequent work broadened this picture in several directions \cite{Bahjat-Abbas:2017htu,CarrilloGonzalez:2019gof,Gurses:2018ckx,Caceres:2025eky}. The Newman--Penrose map associated a class of Kerr--Schild geometries with self-dual Maxwell fields using the optical structure of shear-free null congruences \cite{Elor:2020nqe,Farnsworth:2021wvs}, while the isometry-based Kerr--Schild construction showed that vacuum Kerr--Schild spacetimes possess a distinguished Killing vector whose contraction with the graviton yields the single copy, and that this definition implies the Weyl double copy in the Petrov type--D sector \cite{Luna:2018dpt,Easson:2023dbk}. Recent work has developed a second realization of the Weyl double copy directly in Newman--Penrose language for four-dimensional type--D vacuum spacetimes and extended it to five dimensions \cite{Zhao:2024wtn}. The approach has been extended beyond vacuum solutions,
including the treatment of sources and Einstein--Maxwell systems
\cite{Easson:2020esh,Easson:2021asd,Easson:2022zoh,Armstrong-Williams:2024bog}.

Alongside this modern literature, there exists an older optical route to vacuum Kerr--Schild solutions developed in the early 1970s \cite{Adler:1972ej,Schiffer:1973px}. In the stationary sector, the construction organizes the geometry around the complex expansion of the null congruence. Although historically detached from double-copy ideas, it already contains the ingredients needed to reconstruct the Kerr--Schild congruence from a single complex scalar obeying a Laplace equation together with an eikonal constraint on its inverse. The purpose of this work is to isolate that stationary optical structure and reinterpret it in modern double-copy language.

Our current investigation is deliberately narrow. We focus our attention to stationary vacuum Kerr--Schild spacetimes on a flat background and show that the historical optical ``seed'' construction may be reinterpreted as a local inverse reconstruction from zeroth-copy data to the spacetime geometry. The central object is a complex optical scalar
\begin{equation}
\rho=-\theta-i\omega,
\label{eq:intro_rho}
\end{equation}
where \(\theta\) and \(\omega\) are the expansion and signed twist of the stationary Kerr--Schild congruence.
In this sector, \(\rho\) is harmonic on the flat background, while its inverse \(\mathcal R=-1/\rho\) obeys the eikonal equation and reconstructs the congruence algebraically. In the type--D overlap, the isometry-based Kerr--Schild/Weyl-double-copy analysis implies that the Weyl scalar seed is proportional to the weighted combination \(\rho_{\rm DKS}/P_{\rm DKS}\), where \(P_{\rm DKS}\) is the Debney--Kerr--Schild isometry weight in the unnormalized formalism \cite{Easson:2023dbk}. In the stationary optical normalization adopted here, the seed \(\rho=-\theta-i\omega\) yields the corresponding normalized representative of the zeroth-copy data, while its real part gives the Kerr--Schild profile. We focus here on the expanding stationary Kerr--Schild sector. On regions where \(\rho\) is nonzero, the inverse seed \(\mathcal R \) is well defined and the reconstruction proceeds directly. The nonexpanding sector, including possible Kundt-type branches, is not addressed here and requires a separate analysis.

Our contribution is to place these ingredients in a single spacetime framework adapted to the optical variables \((\rho,\mathcal R,\mathbf{k})\), and to show that the same complex seed organizes the stationary single copy through its gradient. In this restricted sense, the construction clarifies what the zeroth copy encodes in the stationary vacuum Kerr--Schild sector and provides a spacetime realization of part of the geometric structure more commonly organized by the Goldberg--Sachs and twistor frameworks, without invoking twistor methods explicitly.
We do not propose a new general classical double-copy map, nor do we attempt to treat the full non-stationary Kerr--Schild sector. We also do not rederive the full Goldberg--Sachs or Kerr-theorem machinery \cite{2009GReGr..41..433G,Penrose:1987uia,Penrose:1986ca,osti_7216328,1976CMaPh..47...75C}, nor do we replace the broader twistor explanation of the Weyl double copy \cite{White:2020sfn,Chacon:2021wbr,Chacon:2021lox}. Recent work has further clarified the relation between Kerr--Schild and twistor/Penrose-transform double-copy constructions in the self-dual sector~\cite{Albertini:2025ogf}.

The remainder of this paper is organized as follows. In Sec.~\ref{sec:stationary} we fix conventions and derive the stationary optical system, including the harmonic equation for \(\rho\), the eikonal equation for the inverse \(\mathcal R\), and the algebraic reconstruction formula for the Kerr--Schild congruence. In Sec.~\ref{sec:zerothcopy} we show that the same optical seed generates the stationary single copy field strength through its gradient, and we relate this seed to the zeroth-copy data in the type--D overlap, where its real part yields the Kerr--Schild profile. In Sec.~\ref{sec:examples} we illustrate the construction with Schwarzschild and Kerr. We conclude in Sec.~\ref{sec:conclusion} with a discussion of what this stationary optical picture clarifies. 

\section{Stationary optical reconstruction}
\label{sec:stationary}

\subsection{Conventions and setup}
\label{subsec:conventions}

We take the flat background to be four-dimensional Minkowski spacetime in Cartesian coordinates with mostly-minus signature,
\begin{equation}
\eta_{\mu\nu}=\mathrm{diag}(1,-1,-1,-1).
\end{equation}
Greek indices run over spacetime components and Latin indices over spatial components of the background coordinates. Spatial indices are raised and lowered with \(\delta_{ij}\). We use \(\epsilon_{123}=+1\), so that
\begin{equation}
(\nabla\times \mathbf{k})_i=\epsilon_{ijk}\,\partial_j k_k . \nonumber
\end{equation}

Throughout, we restrict attention to stationary vacuum Kerr--Schild spacetimes on a flat background,

\begin{equation}
g_{\mu\nu}=\eta_{\mu\nu}+2V\,k_\mu k_\nu ,
\label{eq:KS_metric}
\end{equation}
where \(k_\mu\) is null with respect to both \(g_{\mu\nu}\) and \(\eta_{\mu\nu}\):
\begin{equation}
g^{\mu\nu}k_\mu k_\nu=0=\eta^{\mu\nu}k_\mu k_\nu .\nonumber
\end{equation}
We choose coordinates adapted to the stationary Killing vector, $X^\mu\partial_\mu=\partial_t$,
and normalize the Kerr--Schild covector so that
$k_0=1$.
The flat-space null condition then implies
$\delta^{ij}k_i k_j=1$,
so the spatial part \(\mathbf{k}\) is a unit vector field.

For comparison with the historical literature \cite{2009GReGr..41.2485K,Bini:2010hrs}, note that one often writes the metric in the form
\begin{equation}
g_{\mu\nu}=\eta_{\mu\nu}-2m\,l_\mu l_\nu,
\qquad
l_\mu=l_0 k_\mu .
\label{eq:historical_KS_form}
\end{equation}
In our notation this is simply a reparameterization of the Kerr--Schild profile,
$V=-m\,l_0^2$. Since the optical equations are for the congruence \(k_\mu\), they are insensitive to whether the scalar prefactor is stored in \(V\) or in \(l_0\).

Rather than defining twist by a square root (and leaving a sign ambiguity) we define the signed optical scalars \(\theta\) and \(\omega\) directly through the decomposition
\begin{equation}
\partial_j k_i=\theta(\delta_{ij}-k_i k_j)+\omega\,\epsilon_{ijl}k_l .
\label{eq:optical_master}
\end{equation}
For stationary vacuum Kerr--Schild congruences, this relation follows from Refs.~\cite{Adler:1972ej,Schiffer:1973px}. We then define the complex optical seed and its inverse by
\begin{equation}
\rho:=-\theta-i\omega,
\qquad
\mathcal R:=-\rho^{-1}.
\label{eq:rho_R_defs}
\end{equation}

\subsection{Optical identities}
\label{subsec:optical_identities}

Equation \eqref{eq:optical_master} immediately implies three basic identities:
\begin{equation}
\nabla\cdot\mathbf{k}=2\theta,
\qquad
\nabla\times\mathbf{k}=-2\omega\,\mathbf{k},
\qquad
(\mathbf{k}\cdot\nabla)\mathbf{k}=0.
\label{eq:basic_optical_identities}
\end{equation}
The first follows by tracing \eqref{eq:optical_master}, the second by taking its curl, and the third by contracting with \(k_j\). Thus \(\theta\) is the expansion and \(\omega\) is the signed twist in our conventions.

The longitudinal derivatives of \(\theta\) and \(\omega\) follow from \eqref{eq:basic_optical_identities}. Taking the divergence of \(\nabla\times\mathbf{k}=-2\omega\,\mathbf{k}\) gives
\begin{equation}
\mathbf{k}\cdot\nabla\omega=-2\theta\omega.\nonumber
\label{eq:kdotomega}
\end{equation}
Likewise, taking the divergence of \((\mathbf{k}\cdot\nabla)\mathbf{k}=0\) yields
\begin{equation}
\mathbf{k}\cdot\nabla\theta=\omega^2-\theta^2.
\label{eq:kdottheta}
\end{equation}

To determine the full gradients of \(\theta\) and \(\omega\), it is convenient to compute \(\nabla^2\mathbf{k}\) in two ways. Differentiating \eqref{eq:optical_master} and using \eqref{eq:basic_optical_identities} gives
\begin{equation}
\nabla^2\mathbf{k}
=
\nabla\theta-(\mathbf{k}\cdot\nabla\theta)\mathbf{k}
+\nabla\omega\times\mathbf{k}
-2(\theta^2+\omega^2)\mathbf{k}.
\label{eq:lapk_first}
\end{equation}
On the other hand, the vector identity
\begin{equation}
\nabla^2\mathbf{k}
=
\nabla(\nabla\cdot\mathbf{k})-\nabla\times(\nabla\times\mathbf{k})\nonumber
\end{equation}
together with \eqref{eq:basic_optical_identities} gives
\begin{equation}
\nabla^2\mathbf{k}
=
2\nabla\theta+2\nabla\omega\times\mathbf{k}-4\omega^2\mathbf{k}.
\label{eq:lapk_second}
\end{equation}
Equating \eqref{eq:lapk_first} and \eqref{eq:lapk_second} and inserting \eqref{eq:kdottheta}, we obtain
\begin{equation}
\nabla\theta=(\omega^2-\theta^2)\mathbf{k}+\mathbf{k}\times\nabla\omega .
\label{eq:grad_theta}
\end{equation}
Taking the cross product of \eqref{eq:grad_theta} with \(\mathbf{k}\) and using \eqref{eq:kdotomega} then gives
\begin{equation}
\nabla\omega=-2\theta\omega\,\mathbf{k}-\mathbf{k}\times\nabla\theta .
\label{eq:grad_omega}
\end{equation}

Combining \eqref{eq:grad_theta} and \eqref{eq:grad_omega} in terms of the complex scalar \(\rho=-\theta-i\omega\) yields the compact equation
\begin{equation}
\nabla\rho=\rho^2\,\mathbf{k}+i\,\nabla\rho\times\mathbf{k}.
\label{eq:rho_master}
\end{equation}
This key stationary optical equation packages the coupled, real equations for the expansion and twist into a single complex relation.

\subsection{Harmonic and eikonal equations}
\label{subsec:harmonic_eikonal}

Equation \eqref{eq:rho_master} immediately implies that \(\rho\) is harmonic on the flat background:
\begin{equation}
\nabla^2\rho=0.
\label{eq:rho_harmonic}
\end{equation}
To see this, take the divergence of \eqref{eq:rho_master}. Using
\begin{equation}
\nabla\cdot(\rho^2\mathbf{k})
=
2\rho\,\nabla\rho\cdot\mathbf{k}
+\rho^2\nabla\cdot\mathbf{k},\nonumber
\end{equation}
together with
\begin{equation}
\nabla\cdot(\nabla\rho\times\mathbf{k})
=
-\nabla\rho\cdot(\nabla\times\mathbf{k}),\nonumber
\end{equation}
and the identities in \eqref{eq:basic_optical_identities}, we find
\begin{equation}
\nabla^2\rho
=
2\rho\,\nabla\rho\cdot\mathbf{k}
+2\theta\rho^2
+2i\omega\,\nabla\rho\cdot\mathbf{k}.    
\label{eq:laprho_intermediate}
\end{equation}
Dotting \eqref{eq:rho_master} with \(\mathbf{k}\) gives
\begin{equation}
\nabla\rho\cdot\mathbf{k}=\rho^2.\nonumber
\label{eq:kdotrho}
\end{equation}
Substituting this into \eqref{eq:laprho_intermediate} yields
\begin{equation}
\nabla^2\rho
=
2\rho^3+2\theta\rho^2+2i\omega\rho^2
=
2\rho^2(\rho+\theta+i\omega)=0,\nonumber
\end{equation}
which proves \eqref{eq:rho_harmonic}.

Using the inverse seed \eqref{eq:rho_R_defs}, and
since \(\nabla\rho=\mathcal R^{-2}\nabla\mathcal R\), equation \eqref{eq:rho_master} becomes
\begin{equation}
\nabla\mathcal R=\mathbf{k}+i\,\nabla\mathcal R\times\mathbf{k}.
\label{eq:R_master}
\end{equation}
We refer to Eq.~\eqref{eq:R_master} as the inverse optical relation, since it is the \(\mathcal R=-1/\rho\) form of the stationary optical equation \eqref{eq:rho_master}.

Dotting \eqref{eq:R_master} with \(\mathbf{k}\) gives
\begin{equation}
\mathbf{k}\cdot\nabla\mathcal R=1.
\label{eq:kdotR}
\end{equation}
Squaring \eqref{eq:R_master} and using \eqref{eq:kdotR} then yields
\begin{equation}
(\nabla\mathcal R)^2=1.
\label{eq:R_eikonal}
\end{equation}
Thus the inverse optical seed satisfies the flat-space eikonal equation.

The eikonal relation \eqref{eq:R_eikonal} suggests a useful
interpretation. In the stationary sector, \(\mathcal R\) may be viewed
as a reduced complex Hamilton--Jacobi, or characteristic, function for the
congruence: its gradient data determine the null directions, and the full
Kerr--Schild vector is reconstructed algebraically from \(\mathcal R\) in
Sec.~\ref{subsec:reconstruction}. In the Schwarzschild case this reduces to the
familiar real eikonal solution \(\mathcal R=r\), while in the Kerr case
\(\mathcal R\) is genuinely complex. In this sense, the zeroth-copy seed
\(\rho\) is naturally related by inversion to a characteristic function for the
congruence.

Equations \eqref{eq:rho_harmonic} and \eqref{eq:R_eikonal} are the central scalar equations of the stationary optical construction. The first demonstrates that the complex seed is harmonic, while the second shows that its inverse has unit gradient norm.

\subsection{Reconstruction of the congruence}
\label{subsec:reconstruction}

The inverse relation \eqref{eq:R_master} determines the Kerr--Schild congruence algebraically. Taking its complex conjugate gives
\begin{equation}
\nabla\mathcal R^*=\mathbf{k}-i\,\nabla\mathcal R^*\times\mathbf{k}.
\label{eq:Rbar_master}
\end{equation}
Define complex spatial vectors \(a:=\nabla\mathcal R\) and \(b:=\nabla\mathcal R^*\).
Then \eqref{eq:R_master} and \eqref{eq:Rbar_master} imply
\begin{equation}
a=\mathbf{k}+i\,a\times\mathbf{k},
\qquad
b=\mathbf{k}-i\,b\times\mathbf{k},\nonumber
\end{equation}
and hence
$a\cdot\mathbf{k}=b\cdot\mathbf{k}=1$ .

Moreover, since \(\mathcal R\) obeys the eikonal equation \eqref{eq:R_eikonal},
\begin{equation}
a\cdot a=(\nabla\mathcal R)^2=1,
\qquad
b\cdot b=(\nabla\mathcal R^*)^2=1.\nonumber
\end{equation}

Now define
\begin{equation}
N:=a+b-i\,a\times b .
\label{eq:N_def}
\end{equation}
Because \(b=a^*\), we have
\begin{equation}
a\cdot b
=
\nabla\mathcal R\cdot\nabla\mathcal R^*
=
\sum_i |\mathcal R_{,i}|^2 .\nonumber
\end{equation}
Moreover, since \(a\cdot a=1\),
\begin{equation}
1=|a\cdot a|
=
\left|\sum_i a_i^2\right|
\le
\sum_i |a_i|^2
=
a\cdot a^*
=
a\cdot b,\nonumber
\end{equation}
so \(a\cdot b\ge 1\), and hence \(1+a\cdot b>0\). We now define
\begin{equation}
\tilde{\mathbf{k}}:=\frac{N}{1+a\cdot b}.
\end{equation}
This vector is real: \(a+b\) is real, while \(a\times b\) is purely imaginary, so
\(-i\,a\times b\) is real.

Using \(a\cdot a=1\), we compute
\begin{align}
(1+a\cdot b)\bigl(\tilde{\mathbf{k}}+i\,a\times\tilde{\mathbf{k}}\bigr)
&=N+i\,a\times N
\nonumber\\
&=a+b-i\,a\times b+i\,a\times\bigl(a+b-i\,a\times b\bigr)
\nonumber\\
&=a+b+a\times(a\times b)
\nonumber\\
&=a+b+a(a\cdot b)-b(a\cdot a)
\nonumber\\
&=a(1+a\cdot b),\nonumber
\end{align}
where we used \(a\times a=0\) and
\[
a\times(a\times b)=a(a\cdot b)-b(a\cdot a).
\]
Therefore
\begin{equation}
a=\tilde{\mathbf{k}}+i\,a\times\tilde{\mathbf{k}}.\nonumber
\end{equation}

Since both \(\tilde{\mathbf{k}}\) and \(\mathbf{k}\) are real and both satisfy
\[
a=x+i\,a\times x,
\]
their difference
\begin{equation}
\delta:=\tilde{\mathbf{k}}-\mathbf{k}
\end{equation}
obeys
$\delta+i\,a\times\delta=0$.

Dotting this with \(\delta\) gives
$\delta^2+i\,\delta\cdot(a\times\delta)=0$.
But \(\delta\cdot(a\times\delta)=0\), so \(\delta^2=0\). Because \(\delta\) is real, it follows that
$\tilde{\mathbf{k}}=\mathbf{k}$.
Hence,
$N=(1+a\cdot b)\,\mathbf{k}$.

Therefore the spatial Kerr--Schild vector is reconstructed by
\begin{equation}
\mathbf{k}
=
\frac{a+b-i\,a\times b}{1+a\cdot b},\nonumber
\end{equation}
or, in components,
\begin{equation}
k_i=
\frac{\mathcal R_{,i}+\mathcal R^*_{,i}
-i\,\epsilon_{ijk}\mathcal R_{,j}\mathcal R^*_{,k}}
{1+\mathcal R_{,l}\mathcal R^*_{,l}}.
\label{eq:k_reconstruction}
\end{equation}

This formula completes the stationary optical reconstruction: once the complex seed
\(\rho\) is known, its inverse \(\mathcal R\) solves the eikonal equation and
determines the congruence algebraically through \eqref{eq:k_reconstruction}. The
stationary vacuum Kerr--Schild geometry is therefore organized by a single complex
optical scalar.

\paragraph*{Local converse in optical variables.}
The above reconstruction may be read in the converse direction.
Let \(\mathcal R\) be a complex function on a region \(U\subset\mathbb R^3\)
such that, away from its zero set,
\[
(\nabla \mathcal R)^2=1,
\qquad
\mathcal R\,\nabla^2\mathcal R=2.
\]
Define \(\rho=-1/\mathcal R\) and reconstruct \(\mathbf{k}\) from \(\mathcal R\)
using Eq.~\eqref{eq:k_reconstruction}. Then \(\mathbf{k}\) is real,
\(\mathbf{k}^2=1\), and \(\rho\) satisfies the optical master relation
\eqref{eq:rho_master}; in particular, \(\rho\) is harmonic.
Thus \((\rho,\mathbf{k})\) furnish precisely the stationary optical data of the
classical generating-function construction \cite{Adler:1972ej,Schiffer:1973px},
now written in the variables used here. By that construction, taking
\(V=m\,\Re\rho\) yields a local stationary vacuum Kerr--Schild metric on the
flat background. In this sense, within the stationary sector, the optical seed
may be viewed as local inverse data for the geometry and the single copy.

\section{Zeroth copy from the optical seed}
\label{sec:zerothcopy}

As we shall see, in the stationary vacuum Kerr--Schild sector, the optical scalar \eqref{eq:intro_rho} does more than reconstruct the null congruence. The Kerr--Schild/Weyl-double-copy analysis of vacuum Kerr--Schild spacetimes gives \cite{Luna:2018dpt,Easson:2023dbk}
\begin{equation}
\Psi_2=\frac{m\rho^3}{P^3},
\qquad
\phi_1=\frac{m\rho^2}{P^2},
\qquad
S=\frac{2m}{3}\,\frac{\rho}{P},
\label{eq:WDC_bridge}
\end{equation}
where \(P:=X\!\cdot\!k\) is the isometry weight associated with the distinguished Killing vector \(X^\mu\), \(\Psi_2\) is the Newman--Penrose Weyl scalar in the type--D sector, \(\phi_1\) is the nonvanishing Maxwell scalar, and \(S\) is the Weyl-double-copy scalar. In the same framework, the real projection of the Weyl-double-copy scalar is
\begin{equation}
\varphi=\frac34\,(S+\tilde S),
\label{eq:varphi_real_projection}
\end{equation}
where in the real Lorentzian stationary sector one has \(\tilde S=\bar S\) \cite{Easson:2023dbk}.

In our stationary normalization,
\begin{equation}
X^\mu\partial_\mu=\partial_t,
\qquad
k_0=1,
\qquad
P=X\!\cdot\!k=1,
\label{eq:stationary_norm_P1}
\end{equation}
so \eqref{eq:WDC_bridge} reduces to
\begin{equation}
S=\frac{2m}{3}\,\rho.
\label{eq:S_equals_rho}
\end{equation}
Thus, in the stationary type--D Kerr--Schild sector, the complex optical seed is a
normalized representative of the Weyl zeroth-copy data \eqref{eq:S_equals_rho}. The harmonicity of \(S\) is then immediate from the harmonicity of \(\rho\).

Using \eqref{eq:S_equals_rho}, the real projection of \(S\) becomes
\begin{equation}
\frac34\,(S+\bar S)=m\,\Re\rho=-m\theta.
\label{eq:real_projection_theta}
\end{equation}
On the other hand, the vacuum Kerr--Schild conditions imply
\begin{equation}
V=\frac{m}{2P^3}\,(\rho+\bar\rho),
\label{eq:V_rho_general}
\end{equation}
so in the same stationary normalization \(P=1\),
\begin{equation}
V=m\,\Re\rho=-m\theta.
\label{eq:V_equals_realrho}
\end{equation}
Hence, in the stationary normalization adopted here,
\begin{equation}
V=\varphi=\frac34\,(S+\bar S).
\label{eq:V_varphi_stationary}
\end{equation}
Equivalently, if one writes the Kerr--Schild metric in the one-function form
\begin{equation}
g_{\mu\nu}=\eta_{\mu\nu}+\phi_{\rm KS}\,k_\mu k_\nu,
\qquad
\phi_{\rm KS}:=2V,
\label{eq:phiKS_def}
\end{equation}
which is the normalization used in the original literature, then
\begin{equation}
\phi_{\rm KS}=\frac32\,(S+\bar S).
\label{eq:phiKS_from_S}
\end{equation}
In the present conventions these quantities are related, in the stationary
normalization \(P=1\), by
\(\varphi=V\) and \(\phi_{\rm KS}=2V=2\varphi\).
Thus the complex object carrying the optical information is \(S\propto \rho\),
while the real Kerr--Schild profile is recovered from its real projection after
the stationary normalization is imposed.
In general, $\varphi$ and $V$ are related by additional factors of the isometry weight, $P$, as seen in \eqref{eq:varphi_real_projection} and \eqref{eq:V_rho_general}.

Equations \eqref{eq:S_equals_rho}--\eqref{eq:phiKS_from_S} show that, in the stationary type--D overlap, the complex optical seed \(\rho\) gives the Weyl zeroth copy, while its real part gives the Kerr--Schild profile. 

\subsection{The stationary single copy from the optical seed}
\label{subsec:singlecopy}

We now show how the same optical seed that reconstructs the stationary Kerr--Schild geometry also
organizes the stationary single copy. In the one-function Kerr--Schild
normalization,
\begin{equation}
g_{\mu\nu}=\eta_{\mu\nu}+\phi_{\rm KS}\,k_\mu k_\nu,
\qquad
\phi_{\rm KS}=2V=2m\,\Re\rho=-2m\theta,\nonumber
\end{equation}
the standard Kerr--Schild single copy is the Maxwell potential~\cite{Monteiro:2014cda}
\begin{equation}
A_\mu=\phi_{\rm KS}\,k_\mu .\nonumber
\end{equation}
In the stationary normalization \(k_0=1\), this gives
\begin{equation}
A_0=\phi_{\rm KS}=-2m\theta,
\qquad
\mathbf A=\phi_{\rm KS}\,\mathbf k=-2m\theta\,\mathbf k.\nonumber
\end{equation}

Defining
\begin{equation}
E_i:=F_{0i},
\qquad
B^i:=\frac12\,\epsilon^{ijk}F_{jk},\nonumber
\end{equation}
with \(F_{\mu\nu}=\partial_\mu A_\nu-\partial_\nu A_\mu\), stationarity implies
\begin{equation}
\mathbf E=2m\,\nabla\theta .\nonumber
\end{equation}
Likewise,
\begin{equation}
\mathbf B=\nabla\times\mathbf A
=-2m\bigl(\nabla\theta\times\mathbf k+\theta\,\nabla\times\mathbf k\bigr).\nonumber
\end{equation}
Using
\begin{equation}
\nabla\times\mathbf k=-2\omega\,\mathbf k,
\qquad
\nabla\omega=-2\theta\omega\,\mathbf k-\mathbf k\times\nabla\theta,\nonumber
\end{equation}
one finds
\begin{equation}
\mathbf B=-2m\,\nabla\omega .\nonumber
\end{equation}
Therefore
\begin{equation}
\mathbf E-i\mathbf B
=
2m\,\nabla\theta+2im\,\nabla\omega
=
-2m\,\nabla\rho,\nonumber
\end{equation}
since \(\rho=-\theta-i\omega\).

In the type--D overlap, where \(S=\frac{2m}{3}\rho\), this becomes
\begin{equation}
\mathbf E-i\mathbf B=-3\,\nabla S.\nonumber
\end{equation}
Thus the same complex optical seed that furnishes the zeroth copy also generates the stationary single copy through its gradient. Its real part governs the Coulombic sector through the expansion, while its imaginary part governs the magnetic-like sector through the twist. The distinction between the purely electric Schwarzschild single copy and the electric-plus-magnetic Kerr single copy is therefore already encoded at the level of the complex optical seed.

\section{Examples}
\label{sec:examples}

Here we illustrate the stationary optical construction in the two canonical vacuum Kerr--Schild examples. In each case the logic is the same: choose the optical seed \(\rho\), verify that \(\rho\) is harmonic and that \(\mathcal R\) obeys the eikonal equation, reconstruct the Kerr--Schild congruence from \(\mathcal R\), and then read off the metric profile from the real part of \(\rho\).

\subsection{Schwarzschild}
\label{subsec:schwarzschild}

For Schwarzschild we take
\begin{equation}
\rho=-\frac{1}{r},
\qquad
\mathcal R=r:=\sqrt{x^2+y^2+z^2}.\nonumber
\end{equation}
(Here \(\mathcal R\) is still the inverse optical seed; in the Schwarzschild case it happens to coincide with the ordinary Euclidean radial coordinate \(r\).)

Away from the origin, \(\rho\) is harmonic, while \(\mathcal R\) satisfies the eikonal equation,
\begin{equation}
(\nabla \mathcal R)^2=(\nabla r)^2=1.\nonumber
\end{equation}
Since \(\mathcal R\) is real, the reconstruction formula \eqref{eq:k_reconstruction} simplifies to
$k_i=\mathcal R_{,i}$,
and therefore
\begin{equation}
\mathbf{k}
=
\nabla r
=
\left(\frac{x}{r},\frac{y}{r},\frac{z}{r}\right).
\label{eq:Sch_k}
\end{equation}
The optical scalars follow immediately from \(\rho=-\theta-i\omega\):
\begin{equation}
\theta=\frac{1}{r},
\qquad
\omega=0.\nonumber
\end{equation}
Equivalently, differentiating \eqref{eq:Sch_k} directly gives
\begin{equation}
\partial_j k_i
=
\frac{\delta_{ij}}{r}-\frac{x_i x_j}{r^3}
=
\frac{1}{r}(\delta_{ij}-k_i k_j),\nonumber
\end{equation}
which is precisely \eqref{eq:optical_master} with \(\omega=0\).

The Kerr--Schild profile is fixed by the real part of the seed,
\begin{equation}
V=m\,\Re\rho=-\frac{m}{r},\nonumber
\end{equation}
and the metric becomes
\begin{equation}
\begin{aligned}
ds^2
&=
dt^2-dx^2-dy^2-dz^2
\\
&\quad
-\frac{2m}{r}
\left(
 dt+\frac{x\,dx+y\,dy+z\,dz}{r}
\right)^2 .
\end{aligned}\nonumber
\end{equation}
Thus the Schwarzschild solution is generated entirely by the real optical seed \(\rho\). In particular, the absence of twist is reflected in the reality of the seed.

\subsection{Kerr}
\label{subsec:kerr}

For Kerr, the optical seed is obtained by the familiar complex shift of the Schwarzschild seed \cite{Newman:1965tw,Schiffer:1973px,Bah:2019sda}:
\begin{equation}
\rho=-\frac{1}{\mathcal R},
\qquad
\mathcal R:=\sqrt{x^2+y^2+(z-ia)^2}.\nonumber
\end{equation}
Differentiating, 
gives
\begin{equation}
\partial_x \mathcal R=\frac{x}{\mathcal R},
\qquad
\partial_y \mathcal R=\frac{y}{\mathcal R},
\qquad
\partial_z \mathcal R=\frac{z-ia}{\mathcal R},\nonumber
\end{equation}
and therefore
\begin{equation}
(\nabla \mathcal R)^2
=
\frac{x^2+y^2+(z-ia)^2}{\mathcal R^2}
=
1.\nonumber
\end{equation}
Away from the singular locus \(\mathcal R=0\), the seed is harmonic,
\begin{equation}
\nabla^2\rho=0.
\nonumber
\end{equation}

It is convenient to introduce the usual real oblate-spheroidal radial variable \(r\), defined implicitly by
\begin{equation}
r^4-r^2(x^2+y^2+z^2-a^2)-a^2 z^2=0,\nonumber
\end{equation}
so that
\begin{equation}
\mathcal R=r-\frac{i a z}{r}.
\nonumber
\end{equation}
Then
\begin{equation}
\rho
=
-\frac{1}{r-i a z/r}
=
-\frac{r^3+i a r z}{r^4+a^2 z^2},\nonumber
\end{equation}
from which we read off
\begin{equation}
\theta=\frac{r^3}{r^4+a^2 z^2},
\qquad
\omega=\frac{a r z}{r^4+a^2 z^2}.\nonumber
\end{equation}
The seed is now genuinely complex: its real part controls the Kerr--Schild profile, while its imaginary part carries the twist of the principal null congruence.

Substituting \(\mathcal R\) into the reconstruction formula \eqref{eq:k_reconstruction} yields the standard Kerr congruence \cite{Kerr:1963ud,Adamo:2014baa},
\begin{equation}
k_x=\frac{r x+a y}{r^2+a^2},
\qquad
k_y=\frac{r y-a x}{r^2+a^2},
\qquad
k_z=\frac{z}{r},\nonumber
\end{equation}
or, equivalently,
\begin{equation}
k_\mu dx^\mu
=
dt
+\frac{r x+a y}{r^2+a^2}\,dx
+\frac{r y-a x}{r^2+a^2}\,dy
+\frac{z}{r}\,dz.\nonumber
\end{equation}
The Kerr--Schild profile is again obtained from the real part of the seed:
\begin{equation}
V=m\,\Re\rho=-\frac{m r^3}{r^4+a^2 z^2}.\nonumber
\end{equation}
The metric therefore takes the form
\begin{equation}
\begin{aligned}
ds^2 ={}& dt^2-dx^2-dy^2-dz^2 \\
&-\frac{2m r^3}{r^4+a^2 z^2}
\Bigg(
 dt
 +\frac{r x+a y}{r^2+a^2}\,dx
 +\frac{r y-a x}{r^2+a^2}\,dy
\\
&\qquad\qquad
 +\frac{z}{r}\,dz
\Bigg)^2 .
\end{aligned}\nonumber
\end{equation}

These examples illustrate the essential difference between the two cases. For Schwarzschild the seed is real and the congruence is twist-free, whereas for Kerr the seed is genuinely complex and its imaginary part directly encodes the twist. In the type--D overlap, the same complex seed is the Weyl zeroth copy up to normalization. In this sense, the familiar Newman--Janis complex shift from Schwarzschild to Kerr is already reflected at the level of the zeroth copy \cite{Janis:1965tx,Newman:1965tw}.

\section{Conclusion}
\label{sec:conclusion}

We have analyzed a historical optical construction for stationary vacuum Kerr--Schild spacetimes and recast it in modern double-copy context. We found the geometry is organized by the complex optical scalar
$\rho$ \eqref{eq:intro_rho}
built from the expansion and signed twist of the Kerr--Schild congruence. We showed that \(\rho\) is harmonic on the flat background, that its inverse \(\mathcal R=-1/\rho\) obeys the eikonal equation, and that \(\mathcal R\) reconstructs the Kerr--Schild congruence algebraically.

In the stationary type--D overlap, this complex optical seed furnishes a normalized representative of the Weyl zeroth-copy data, while its real part yields the Kerr--Schild profile and its gradient generates the stationary single-copy field strength. The spacetime, corresponding zeroth-copy data, and associated gauge field are therefore all organized by a single complex optical quantity. In this sense, the construction clarifies what the zeroth copy encodes in the stationary vacuum Kerr--Schild sector and provides a spacetime realization of part of the geometric structure, without invoking twistor methods explicitly.

Schwarzschild and Kerr make this concrete: the seed is real in the twist-free Schwarzschild case, but genuinely complex for Kerr, where its imaginary part encodes the twist of the principal null congruence. Thus the complex optical seed carries geometric information beyond the real Kerr--Schild profile alone. A useful conceptual lesson emerges from these two canonical examples. The
singularity of the vacuum optical seed is not accidental. In the stationary
asymptotically flat vacuum sector, \(\rho\) is harmonic on the auxiliary flat
background away from its singular set and decays at spatial infinity. A
nontrivial decaying harmonic seed on flat \(\mathbb R^3\) therefore cannot be
globally smooth everywhere, so singular seed behavior is the flat-background
imprint of nontrivial conserved gravitational data such as mass and angular momentum. In this sense, the
Schwarzschild monopole seed and the complex-shifted Kerr seed realize the same
basic obstruction in different forms.

Our claims have been intentionally limited. We have not treated the full non-stationary Kerr--Schild sector, nor have we proposed a new general classical double-copy map. Rather, the present paper is intended to clarify what the zeroth copy encodes in the stationary vacuum Kerr--Schild sector and provides a bridge between optical solution-generating methods and modern double-copy constructions.

A natural next step is to extend the seed viewpoint beyond a flat background.
For the Schwarzschild--(A)dS family one may write
\(
g_{\mu\nu}=\bar g^{(\Lambda)}_{\mu\nu}+2V\,k_\mu k_\nu
\),
with \(\bar g^{(\Lambda)}_{\mu\nu}\) a maximally symmetric background satisfying
\(\bar R=4\Lambda\). In that setting, the flat-background seed equations appear
to deform in a simple curvature-dependent way: the radial seed is expected to obey a conformally coupled scalar equation rather
than the flat-space Laplace equation. This suggests that the optical-seed
picture may persist in the simplest constant-curvature example, although a full
curved-background reconstruction, including the analog of the inverse-seed
relation, lies beyond the scope of the present work.

More broadly, it would be interesting to understand whether a comparably sharp
inverse interpretation survives in wider classes of algebraically special
geometries, whether the optical seed can be extended usefully beyond the
type--D overlap, and whether a controlled non-stationary generalization can be
formulated without losing the geometric transparency found here. Even within the stationary sector, the optical viewpoint suggests that the relevant zeroth-copy data are best understood not merely as an auxiliary scalar, but as a compact encoding of congruence structure -- equivalently, through the inverse seed, of the characteristic data from which both the spacetime and the corresponding gauge field emerge.
\vspace{.6cm}

\acknowledgments
It is a pleasure to thank Gabe Herczeg, Cindy Keeler, Tucker Manton and Max Pezzelle for useful conversations. 
This work is supported, in part, by the U.S. Department of Energy, Office of High Energy Physics, under Award Number DE-SC0019470.

\bibliography{bib}

\end{document}